\documentclass[fleqn,usenatbib]{mnras}

\usepackage{newtxtext,newtxmath}
\usepackage{lscape}
\usepackage[T1]{fontenc}

\DeclareRobustCommand{\VAN}[3]{#2}
\let\VANthebibliography\thebibliography
\def\thebibliography{\DeclareRobustCommand{\VAN}[3]{##3}\VANthebibliography}

\usepackage{graphicx}	% Including figure files
\usepackage{amsmath}	% Advanced maths commands
\usepackage{pdflscape}
\title[Meteorite dropper spring 2022]{Identifying meteorite droppers among the population of bright 'sporadic' bolides imaged by the Spanish Meteor Network during the spring of 2022}

\author[E. Peña-Asensio et al.]{
E. Peña-Asensio,$^{1,2}$\thanks{E-mail: eloy.pena@uab.cat, eloy.peas@gmail.com}
J. M. Trigo-Rodríguez,$^{2,3}$
A. Rimola,$^{1}$
M. Corretgé-Gilart,$^{4}$
and D. Koschny$^{5}$
\\
$^{1}$Departament de Química, Universitat Autònoma de Barcelona 
08193 Bellaterra, Catalonia, Spain\\
$^{2}$Institut de Ciències de l’Espai (ICE, CSIC),
Campus UAB, C/ de Can Magrans s/n, 
08193 Cerdanyola del Vallès, Catalonia, Spain\\
$^{3}$Institut  d’Estudis  Espacials  de  Catalunya  (IEEC),
08034  Barcelona,  Catalonia,  Spain\\
$^{4}$Universitat Politècnica de Catalunya (UPC), Carrer de Jordi Girona, 31, 08034 Barcelona, Spain\\
$^{5}$TU Munich, Boltzmannstrasse 15, 85748 Garching, Germany
}

\date{Accepted XXX. Received YYY; in original form ZZZ}

\pubyear{2022}

\begin{document}
\label{firstpage}
\pagerange{\pageref{firstpage}--\pageref{lastpage}}
\maketitle

\begin{abstract}
The extraordinary weather conditions available between February and March 2022 over Spain have allowed us to analyze the brightest fireballs recorded by the monitoring stations of the Spanish Meteor Network (SPMN). We study the atmospheric flight of 15 large meteoroids to determine if they are meteorite dropper events to prepare campaigns to search for freshly fallen extraterrestrial material. We investigate their origins in the Solar System and their dynamic association with parent bodies and meteoroid streams. Employing our Python pipeline \textit{3D-FireTOC}, we reconstruct the atmospheric trajectory utilizing ground-based multi-station observations and compute the heliocentric orbit. In addition, we apply an ablation model to estimate the initial and terminal mass of each event. Using a dissimilarity criterion and propagating backward in time, we check the connection of these meteoroids with known complexes and near-Earth objects. We also calculate if the orbits are compatible with recent meteoroid ejections. We find that $\sim$27\% of these fireballs are dynamically associated with minor meteoroid streams and exhibit physical properties of cometary bodies, as well as one associated with a near-Earth asteroid. We identify two meteorite-producing events; however, the on-site search was unsuccessful. By considering that these fireballs are mostly produced by cm-sized rocks that might be the fragmentation product of much larger meteoroids, our findings emphasize the idea that the population of near-Earth objects is a source of near-term impact hazards, existing large Earth-colliding meteoroids in the known complexes.
\end{abstract}

\begin{keywords}
meteorites, meteors, meteoroids --
                comets: general --
                minor planets, asteroids: general
\end{keywords}

%%%%%%%%%%%%%%%%%%%%%%%%%%%%%%%%%%%%%%%%%%%%%%%%%%

%%%%%%%%%%%%%%%%% BODY OF PAPER %%%%%%%%%%%%%%%%%%

\section{Introduction}
%--------------------------------------------------------------------
The interplanetary medium is composed of countless millimeter- and centimeter-sized objects called meteoroids, some of which eventually cross the path of our planet \citep{Brown2002, Murad2002, Trigo2022}. These small bodies are fragments produced by the catastrophic disruption or collisions of comets, asteroids, or even impacts on planets \citep{Chapman2010, Toth2011, Gritsevich2012, Trigo2014}. Due to tidal forces and sublimation by high temperatures of the Sun, cometary aggregates and rubble pile asteroids with efficient disruption processes suffer fragmentations in their passage through the perihelion, scattering meteoroids throughout their orbit that constitute the so-called meteoroid streams (also known as meteor showers) \citep{Jenniskens1994, Jenniskens1998, Jenniskens2006, Vaubaillon2019}. Some of these meteoroid streams have Earth-intersecting orbits, so they are generally repeated in annual cycles. After experiencing different physical phenomena such as orbital perturbations, impacts with other objects, Yarkovsky, YORP, or Poynting-Robertson effect, other meteoroids suffer time scale decoherence and end up their space travel impacting on our planet as sporadic events, that is, apparently not associated with any known complex \citep{Olsson1986, Bottke2000, PaulsGladman2005, Broz2006, Koschny2019}. 

The impact of these objects at high velocity with the upper part of our atmosphere produces a luminous phase in the visible range due to the collision with the atoms of the air and the consequent melting, evaporation, and progressive ionization of the meteoroid material \citep{Ceplecha1998, Silber2018AdSpR}. This phenomenon is known as a meteor and is called a fireball or bolide if its magnitude is greater than that of Venus. From the observation and analysis of fireballs with ground-based multi-stations, more than 10 major showers have been established (Quadrantids, April Lyrids, $\eta$-Aquarids, Southern $\delta$-Aquariids, Perseids, Orionids, Taurids, Leonids, Geminids and Ursids), that is, meteoroid streams that present activity of more than 10-15 meteors per hour \citep{Bagnall2021}. However, there are hundreds of minor showers with lower activities as well as near-Earth asteroids, many of them poorly studied, that can produce bright fireballs and, therefore, potentially meteorite dropper events, just as being a source of impact hazard to the Earth \citep{Voloshchuk1996, Halliday1987, Madiedo2008, Borovi2015, Trigo2017, TrigoBlum2022MNRAS, PeAs2022}.

The months between January and April are especially relevant from the meteor science point of view as meteorite fall rates display a peak during the beginning of spring in either hemisphere \citep{Halliday1982}. Unfortunately, the weather during winter and spring is usually not helpful for fireball monitoring and clouds generally prevent detailed trajectory reconstruction and strewn-field estimates. In this sense, the months of February and March 2022 were especially clement in the Spanish territory so the Spanish Meteor Network (SPMN) has been able to record and analyze several spectacular fireballs, many of them associated with minor meteoroid streams rather than being sporadic.

In section \ref{secDataMethod}, we first outline the SPMN network's current infrastructure that has allowed recording these events with multiple stations. We also mention the methodology applied for fireball analysis. In section \ref{secAtmHel}, we describe the results of the atmospheric flight reconstruction, terminal mass prediction, and heliocentric orbit calculation. In section \ref{secDynAsso}, we analyze the dynamic associations with parent bodies, near-Earth asteroids and comets, and minor and major meteoroid streams. In addition, we examined the compatibility of these events being recently ejected meteoroids. Finally, we discuss the results in section \ref{secDisc} and offer our conclusions in section \ref{secConcl}.

%--------------------------------------------------------------------
\section{Data collection and methodology}\label{secDataMethod}
%--------------------------------------------------------------------
Since its creation in 2005, thanks to the operability of the SPMN network, the whole sky of continental Spain is monitored full time, the last decade also including the Balearic and Canary Islands. Currently, a total of 34 stations with charged-coupled device (CCD) video and all-sky cameras are operational, some of them equipped with spectrometers. In addition, three forward-scatter detectors monitor radio meteors \citep{Trigo2004}. The stations involved in the events analyzed in this work are shown in Table \ref{refStations}, also incorporating the recently installed \textit{AllSky7} camera at European Space Agency Cebreros' station. This camera array allowed us to record 169 bright meteors up to an apparent magnitude of -6 between February and March of 2022, from which we selected the 15 largest multi-station bolides for analysis.

\begin{table}
\caption[Stations]{Location of the fireball observation points involved in this work.}
\label{refStations}
\centering
\begin{tabular}{ccccc}
\hline
Station & Name & Long ($^\circ$) & Lat ($^\circ$) & Alt (m) \\
\hline
A & Alpicat & 0.5568 & 41.6676 & 252 \\
B & Barx & -0.3041 & 39.0146 & 336 \\
C & Benicàssim & 0.0386 & 40.0342 & 15 \\
D & Calar Alto & -2.549 & 37.2212 & 2152 \\
E & Cebreros & -4.3693 & 40.4541 & 700 \\
F & Corbera & 1.8906 & 41.4092 & 501 \\
G & Estepa & -4.8766 & 37.2914 & 537 \\
H & GranTeCan & -17.8919 & 28.7567 & 2267 \\
I & La Murta & -1.6756 & 38.0967 & 469 \\
J & Monfragüe & -6.0108 & 39.7736 & 411 \\
K & Morata de Jalón & -1.4821 & 41.474 & 415 \\
L & Olocau & -0.5363 & 39.6744 & 225 \\
M & Playa Blanca & -13.8241 & 28.8747 & 10 \\
N & Puertollano & -4.1129 & 38.7032 & 697 \\
O & Sant Mateu & 0.1758 & 40.465 & 349 \\
 \hline
\end{tabular}
\end{table}

New video processing and trajectory calculation techniques allow the automation of the analysis process of meteors, bolides, and artificial fireballs produced by atmospheric re-entries of human-made objects. We developed the \textit{3D-FireTOC} Python code that automates this study allowing the reconstruction of atmospheric trajectories and the calculation of heliocentric orbits from multiple recordings by using the intersection of planes method \citep{PeAs2021a, PeAs2021b}. Unlike traditional analytical methods, which solve the orbit by correcting for zenith attraction and diurnal aberration \citep{Ceplecha1987}, we have now implemented the accurate IAS15 high-order N-body integrator with an adaptive time step included in the REBOUND package to compute the heliocentric orbit \citep{ReinSpiegel2015}. The integrator is based on the RADAU-15 developed in \citet{Everhart1985} and has a high performance resolving close encounters. We account for the Earth’s and Moon’s oblateness by including the J2 and J4 gravitational harmonic coefficients thanks to the REBOUNDx module \citep{Tamayo2020}.

For most cases, we performed the astrometric calibration by solving the polynomial modification of \citet{Borovi1992} proposed by \citet{Bannister2013}, which exhibits a better convergence while ensuring a very excellent level of uncertainty. To achieve the best fit, we use a simplicial homology global optimization algorithm to find the absolute minimum \citep{Endres2018}. For recordings with sufficient background stars, we apply the method proposed in \citet{Borovi1995}, which produces even lower errors down to 0.01$^\circ$ for azimuth and elevation. All calibrations are also cross-checked with the quadratic model described in \citet{PeAs2021a}.

With the mean uncertainties obtained in the astrometry for the camera calibration fit, we generate 1,000 clones to perform a Monte Carlo simulation following a Gaussian distribution applied to each detected point. We propagate every clone backward starting with its pre-atmospheric velocity from the beginning of the detected luminous phase until they are outside the Earth's influence, specifically, at 10 times the Earth Hill sphere. We then integrate forward to the date of impact but without taking into account the gravitational attraction of the Earth-Moon system to obtain the osculating orbital elements at the time of the detection (referred to the J2000 equinox).

We further perform a backward integration over 10,000 years evaluating the evolution of an orbital dissimilarity criterion to test the dynamic association with parent body candidates. This is necessary as the most favorable candidate at the time of impact is not always the most reliable because it may be the result of a coincidence at that precise date. The meteoroid is integrated with its corresponding 1,000 clones generated from the uncertainties and the meteoroid streams are modeled by 18 equally spaced distributed particles over the true anomaly. Based on the orbital dissimilarity criterion, we assume that an association is robust enough if it remains below the cutoff for 5,000 years, minimizing the probability of being a random association \citep{Porub2004}.

Different techniques have been developed and discussed to establish the association between meteors and meteor showers or parent bodies, and they are still a source of debate today. One of the most established and widely used criteria is $D_{D}$ \citep{Drummond1981}, which is a semi-quantitative approach to measure the dissimilarity of two orbits as a function of their orbital parameters in the five-dimensional phase.

Based on the $D_{SH}$ criterion \citep{SH1963}, the $D_D$ criterion was defined as:
\begin{equation}
\begin{split}
D_{D}^{2}=\left(\frac{e_{B}-e_{A}}{e_{B}+e_{A}}\right)^{2}+\left(\frac{q_{B}-q_{A}}{q_{B}+q_{A}}\right)^{2}+\left(\frac{I_{B A}}{\pi}\right)^{2} + \\
+\left(\frac{e_{B}+e_{A}}{2}\right)^{2}\left(\frac{\theta_{B A}}{\pi}\right)^{2},
	\label{eq:D_D}
\end{split}
\end{equation}

where $e$ is the eccentricity, $q$ is the perihelion distance, $I_{BA}$ is the angle between the orbital planes, $\pi_{BA}$ is the difference between longitudes of perihelia measured from the intersection of both orbits, and $\theta_{BA}$ is the orbit angle between the lines of apsides.

The thresholds of the dissimilarity functions, far from defining an exact barrier, offer an approximation with fair statistical significance, which, in addition, may vary depending on the inclination of the orbits and the population size. Therefore, they are not a defining indicator, and it is also necessary to verify that the orbits are not only similar at a given time but also that this similarity lasts over time. In this sense, we use 0.18 as a cut-off for $D_D$ \citep{Galligan2001}. Although this threshold value is high, we use it as a first filter, but not as the only association condition as we also check its evolution over time.

In addition, we evaluate if the separation of the meteoroid from its possible parent body could have occurred in relatively short timescales. For this purpose, during the orbital integration, we monitor the minimum distance between the objects and the change in the velocity vector that would be needed to move from one orbit to the other one. In this way, we can observe if the velocity change is compatible with typical collisional ejection processes between small bodies.

We also examined Tisserand's parameter with respect to Jupiter $T_j$, which is helpful to determine the evolution of small bodies since it remains broadly constant for long periods. It is used to classify planet-crossing objects, usually, as Jupiter-family comets (JFCs) if $2<T_j<3$ and asteroidal when $T_j>3$.

We evaluate the catastrophic disruption for each event by obtaining the ram pressure at peak brightness, that is, the bulk aerodynamic strength  ($s=\rho\cdot v^2$) accordingly to the U.S. standard atmosphere 1976 \citep{Bronshten1981}. This parameter is typically used to mechanically characterize the meteoroid and to classify the material regarding the bulk density. For events that do not present an explosion, we evaluate the peak of maximum brightness, thus obtaining only an estimate of the lower limit for the composition.

Additionally, assuming an isothermal atmosphere and applying the dynamic third-order time-dependent system for characterizing meteor deceleration based on the velocity ($v$) and the height ($h$), we compute the ballistic coefficient ($\alpha$) and mass loss parameter ($\beta$) \citep{Gritsevich2006SoSyR, Gritsevich2008, Gritsevich2009, Gritsevich2012, Turchak2014JTAM}: 

\begin{equation}
F_i(h_i,v_i,\alpha,\beta) =  2\alpha e^{-h_i} - \Delta _i e^{-\beta},\label{eq:alphabeta}
\end{equation}

with $\Delta _i = \overline{Ei}(\beta) - \overline{Ei}(\beta v_i^2)$, $i = 1, 2, ..., n$, where
\begin{equation}
\overline{Ei}(x) =  \int_{-\infty}^{x} \frac{e^tdt}{t} dx. \nonumber
\end{equation}

These adimensional parameters are defined as

\begin{equation} 
\alpha = \frac{1}{2}c_d \frac{\rho_0h_0S_{0}}{M_{0}\sin{\gamma}},
    \label{alpha}
\end{equation}

and 
\begin{equation}
\beta = (1-\mu)\frac{c_h v_{0}^2}{2c_dH^*},
    \label{beta}
\end{equation}

where $c_d$ is the drag coefficient, $\rho_0$ is the atmospheric density at sea level, $h_0$ is the scale height for a homogeneous atmosphere and $\gamma$ is the slope of the fireball to the local horizon, $M_{0}$ is the meteoroid mass before impacting the top of the atmosphere, $\mu$ is the dimensionless shape change parameter, $c_h$ is the heat transfer coefficient, $v_0$ is the entry velocity, and $H^*$ is the sublimation heat. $\mu$ is a constant value that relates the cross-sectional area $S$ with the mass as follows: $S/S_{0} = (M/M_{0})^\mu$ \citep{Lyytinen2016}. Note that as it is an atmospheric flight dynamics model with an asymptotic solution, the minimization problem itself yields an initial velocity at infinity that corresponds to the pre-atmospheric velocity.

These parameters allow properly describing the atmospheric flight and estimating the meteor fate based on the so-called $\alpha-\beta$ criterion \citep{Sansom2019}. The boundaries that delimit the fall likelihood (with a terminal mass threshold of 50 g) are determined by the two extreme values of the shape change coefficient: $\mu=0$ when the meteoroid is not spinning and $\mu=2/3$ when the meteoroid surface is equally ablated due to the rotation.

From the aerodynamic strength values, we assign a meteoroid bulk density based on \citet{Chyba1993}: cometary if $s<10^5\,Pa$; carbonaceous if  $10^5\,Pa<s<10^6\,Pa$; rocky if \linebreak$10^6\,Pa<s<10^7\,Pa$; and rocky-iron if its aerodynamic strength is greater than $10^7\,Pa$. This allows us to fit the object size $D$, the pre-atmospheric mass $M_{0}$, and the terminal mass $M_{t}$ (the final mass at the end of the luminous atmospheric phase), being

\begin{equation}
M_{0}=\left(\frac{1}{2} \frac{c_d A_{0} \rho_0 h_0}{\alpha \rho_m^{2/3} \sin\gamma}\right)^{3},
	\label{eq:mass_init}
\end{equation}

where $A_{0}$ is the pre-atmospheric shape coefficient.

The terminal mass can be computed using the last observed velocity in the following instant mass equation
\begin{equation}
M(t)=M_{0} e^{-\frac{\beta}{1-\mu} \left(1-\left(\frac{v(t)}{v_0}\right)^2 \right)},
	\label{eq:mass_instant}
\end{equation}

where $v(t)$ is the instantaneous velocity.

%--------------------------------------------------------------------
%--------------------------------------------------------------------

%--------------------------------------------------------------------
\section{Atmospheric flight and heliocentric orbit}\label{secAtmHel}

Once the most suitable recordings of each event have been selected, and the lenses of each camera have been calibrated to correct distortions and found the transformation between pixel and position in the sky, we can apply the triangulation using the weighted method of the intersection of planes for multiple stations to obtain the real position of the meteoroid in each frame. Each station recorded the events in a single shot, except for the grazing meteoroid SPMN080322, which moved out of the field of view. Therefore, we had to combine the recordings from two cameras to obtain the complete luminous trail. Figure \ref{Fireball_comp} shows a composite of overlapping images of some of the events recorded and analyzed in the following section.

In some images, like the one of the SPMN060222 fireball captured in color from Corbera, an intense reddish tone due to the glowing ionized air can be seen, although further color calibrations are necessary for a precise determination of the tone. In the trace drawn during the atmospheric flights, it can be seen how several of them show multiple brightness peaks, as a result of the rapid rotation and differentiated ablation, while others only exhibited a large final flare due to the catastrophic disruption. The beginning and ending position, distance flight, and direction of the luminous phase for each event are shown in Table \ref{AtmTray}. The initial heights range from $\sim$ 120 to 83 km and terminal heights (before starting the dark flight) range from $\sim$ 80 to 13 km. As expected, the azimuth and slope have a random distribution, with the average slope being around 45$^\circ$. Note that the slope is measured with respect to the local horizon, 0$^\circ$ corresponding to a fully grazing meteor. In this regard, we see how the event SPMN010322A traveled through the atmosphere a notably greater distance than the rest ($\sim$198 km), its slope being close to 10$^\circ$. Event SPMN080322A, although also with a shallow slope, underwent a rapid disruption at 70 km altitude, which did not allow it to cover a long distance.

\begin{landscape}

   \begin{figure}
   \centering
   \includegraphics[scale=1.3]{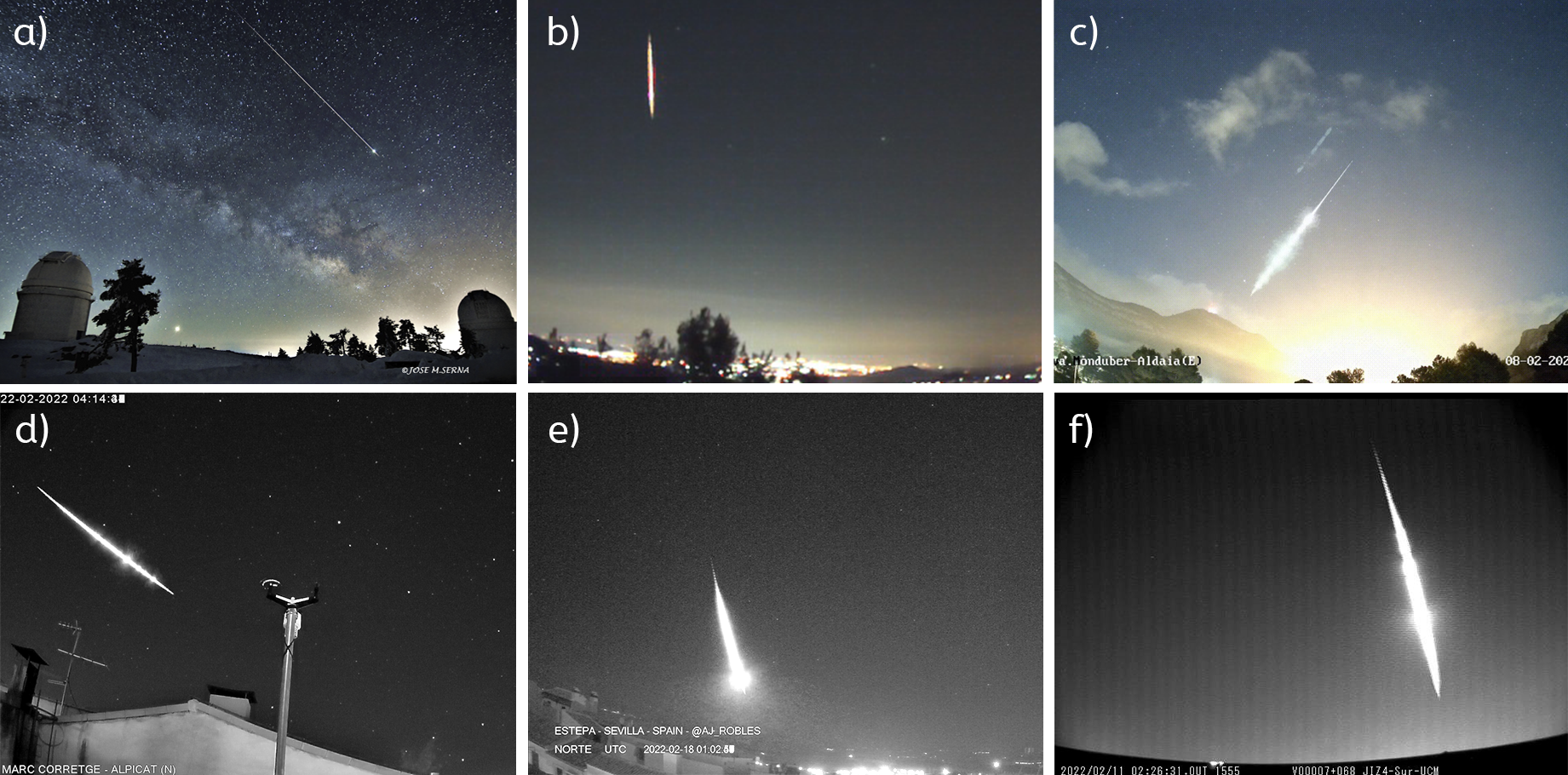}
   \caption{Selection of blended frames of some of the events analyzed in this work: a) SPMN090322C from Calar Alto by José M. Serna García, b) SPMN060222 from Corbera, c) SPMN080222B from Barx, d) SPMN220222 from Alpicat, e) SPMN180222 from Estepa, and f) SPMN110222 from Madrid.}
              \label{Fireball_comp}%
    \end{figure}

\begin{table}
\caption{Recorded fireballs with the beginning and ending position, flight distance traveled, and direction of the atmospheric flight.}\label{AtmTray}
\centering
\scriptsize
 \begin{tabular}{lccccccccccc}
\hline
SPMN code & Datetime (UTC) & Stations & Long$_{0}$ ($^\circ$) & Lat$_{0}$ ($^\circ$) & h$_{0}$ (km) & Long$_{t}$ ($^\circ$) & Lat$_{t}$ ($^\circ$) & h$_{t}$ (km) & Distance (km) & Azimuth ($^\circ$) & Slope ($^\circ$) \\
\hline \hline
060222 & 2022-02-06 23:03:20 & A,F & 4.324$\pm$0.011 & 42.848$\pm$0.004 & 91.3$\pm$0.4 & 4.392$\pm$0.009 & 42.8570$\pm$0.0031 & 69.16$\pm$0.26 & 22.9$\pm$0.5 & 80$\pm$5 & 75$\pm$4 \\
080222A & 2022-02-08 01:09:54 & A,B,K & -2.5529$\pm$0.0029 & 41.2470$\pm$0.0012 & 101.594$\pm$0.028 & -2.5542$\pm$0.0029 & 41.6429$\pm$0.0014 & 41.06$\pm$0.08 & 77.54$\pm$0.32 & 359.4$\pm$0.5 & 51.33$\pm$0.11 \\
080222B & 2022-02-08 23:31:00 & A,B & 1.1353$\pm$0.0010 & 38.9446$\pm$0.0008 & 89.16$\pm$0.07 & 1.1055$\pm$0.0009 & 39.1885$\pm$0.0007 & 36.134$\pm$0.024 & 60.68$\pm$0.22 & 353.97$\pm$0.32 & 60.940$\pm$0.032 \\
110222 & 2022-02-11 02:26:30 & B,E,O & -3.625$\pm$0.009 & 39.702$\pm$0.007 & 89.8$\pm$0.8 & -3.731$\pm$0.005 & 39.455$\pm$0.006 & 37.50$\pm$0.21 & 65.5$\pm$0.5 & 198.7$\pm$2.8 & 52.9$\pm$1.2 \\
140222B & 2022-02-14 20:59:07 & G,I,N & -3.5864$\pm$0.0014 & 37.8739$\pm$0.0004 & 94.646$\pm$0.025 & -3.2628$\pm$0.0009 & 37.78175$\pm$0.00030 & 49.547$\pm$0.018 & 60.51$\pm$0.22 & 109.69$\pm$0.05 & 48.18$\pm$0.06 \\
180222 & 2022-02-18 01:02:45 & I,J,O & -6.1642$\pm$0.0023 & 39.380$\pm$0.004 & 88.79$\pm$0.06 & -6.0776$\pm$0.0034 & 39.5080$\pm$0.0034 & 12.87$\pm$0.15 & 82.9$\pm$0.8 & 26.8$\pm$1.3 & 66.43$\pm$0.18 \\
220222 & 2022-02-22 04:34:24 & A,K & -0.5435$\pm$0.0010 & 42.3780$\pm$0.0005 & 83.77$\pm$0.08 & 0.1736$\pm$0.0004 & 42.2556$\pm$0.0005 & 38.92$\pm$0.04 & 80.57$\pm$0.13 & 102.42$\pm$0.08 & 33.814$\pm$0.015 \\
010322A & 2022-03-01 00:48:01 & A,C,L & 2.6121$\pm$0.0034 & 41.3954$\pm$0.0020 & 95.79$\pm$0.07 & 1.4258$\pm$0.0018 & 39.9335$\pm$0.0015 & 50.499$\pm$0.024 & 197.0$\pm$0.4 & 211.99$\pm$0.07 & 13.293$\pm$0.024 \\
010322B & 2022-03-01 01:43:57 & B,O & -2.793$\pm$0.008 & 39.9817$\pm$0.0019 & 101.07$\pm$0.30 & -3.258$\pm$0.010 & 39.5159$\pm$0.0019 & 71.70$\pm$0.24 & 74.2$\pm$0.7 & 217.7$\pm$1.0 & 23.3$\pm$0.5 \\
080322A & 2022-03-08 00:36:59 & A,F & 0.8633$\pm$0.0008 & 40.6421$\pm$0.0005 & 96.82$\pm$0.08 & 1.7423$\pm$0.0006 & 41.00590$\pm$0.00030 & 80.13$\pm$0.05 & 87.00$\pm$0.12 & 60.904$\pm$0.032 & 11.06$\pm$0.06 \\
080322B & 2022-03-08 19:26:22 & A,L & 1.8383$\pm$0.0005 & 40.4211$\pm$0.0005 & 83.58$\pm$0.05 & 1.8210$\pm$0.0005 & 40.4374$\pm$0.0005 & 36.786$\pm$0.021 & 57.70$\pm$0.18 & 320.613$\pm$0.024 & 54.09$\pm$0.10 \\
090322B & 2022-03-09 03:01:46 & A,B,C & -1.5107$\pm$0.0010 & 39.8088$\pm$0.0004 & 120.65$\pm$0.07 & -2.0243$\pm$0.0011 & 39.94857$\pm$0.00035 & 77.071$\pm$0.035 & 71.19$\pm$0.13 & 289.42$\pm$0.04 & 37.75$\pm$0.12 \\
090322C & 2022-03-09 04:25:38 & D,I & -2.0192$\pm$0.0007 & 36.9452$\pm$0.0009 & 92.94$\pm$0.15 & -2.1597$\pm$0.0006 & 36.4849$\pm$0.0017 & 58.54$\pm$0.10 & 64.80$\pm$0.05 & 193.73$\pm$0.15 & 32.07$\pm$0.23 \\
100322 & 2022-03-10 01:38:19 & H,M & -15.540$\pm$0.014 & 30.0550$\pm$0.0034 & 85.4$\pm$0.8 & -15.600$\pm$0.022 & 29.689$\pm$0.005 & 29.2$\pm$0.6 & 82.5$\pm$0.6 & 188$\pm$5 & 42.94$\pm$0.17 \\
120322 & 2022-03-12 22:15:53 & A,L & 1.1473$\pm$0.0004 & 40.7151$\pm$0.0007 & 94.21$\pm$0.09 & 1.09818$\pm$0.00035 & 40.7597$\pm$0.0006 & 67.85$\pm$0.05 & 27.40$\pm$0.16 & 319.4$\pm$0.4 & 74.30$\pm$0.25 \\
 \hline
 \end{tabular}
\end{table}
\end{landscape}

\begin{figure}
   \centering
   \includegraphics[width=\columnwidth]{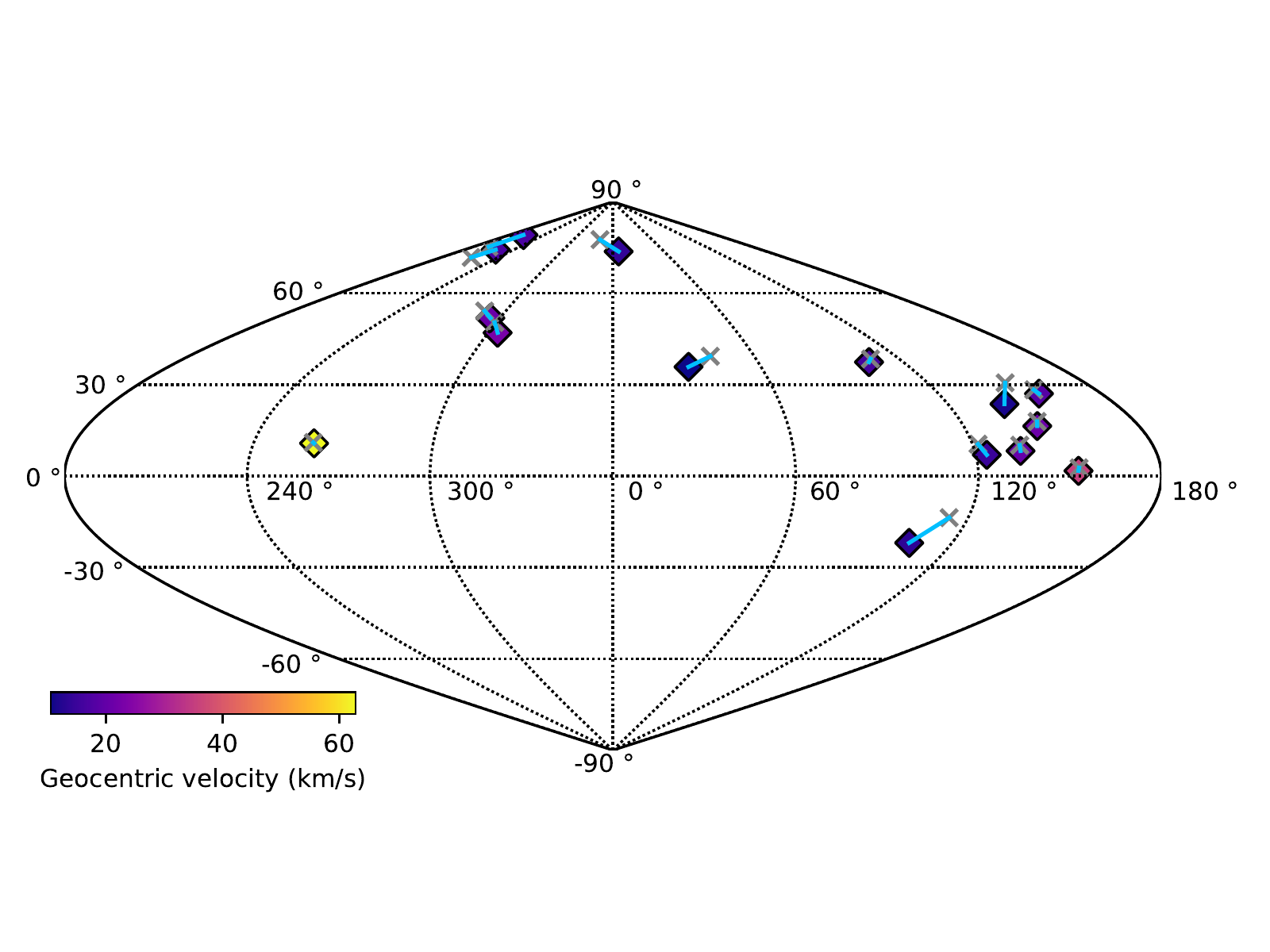}
   \caption{Sinusoidal projection of the geocentric (diamond) and apparent (gray cross) radiants. Radiant pairs are connected with a light blue line. Geocentric radiants are color-coded according to their geocentric velocity.}
              \label{radiants}%
    \end{figure}

Using the height at which the brightest flare occurs, the air density, and the velocity at that point, we calculate the aerodynamic strength. According to the value of this dynamic pressure, we estimate the bulk density as explained in Section \ref{secDataMethod}, which is used to calculate the pre-atmospheric diameter assuming a perfect sphere. To obtain the ballistic coefficient and the mass loss parameter, we assume an aerodynamic drag coefficient of 1.3 and a shape change coefficient of 2/3 \citep{GritsevichKoschny2011}. The geocentric velocities range from $\sim$ 63 to 11 km/s, and most of the radiants are in the northern hemisphere, as depicted in Figure \ref{radiants} in sinusoidal projection. All the computed parameters are shown in Table \ref{AtmAbla} and \ref{RadVel}.

\begin{table*}
\caption{Recorded fireballs with aerodynamic strength, ballistic coefficient, mass loss parameter, pre-atmospheric diameter, pre-atmospheric mass, and terminal mass.}\label{AtmAbla}
\centering
\footnotesize
\begin{tabular}{lcccccccccccc}
\hline
SPMN code & s (kPa) & $\alpha$ & $\beta$ & D (cm) & M$_{0}$ (g) & M$_{t}$ (g) \\
\hline
060222 & 18.9$\pm$0.4 & (8.6$\pm$0.7)$\cdot$10$^2$ & 10.6$\pm$1.0 & 1.17$\pm$0.08 & 0.83$\pm$0.16 & <1 & \\
080222A & 724$\pm$7 & 195.1$\pm$3.4 & 1.023$\pm$0.031 & 6.35$\pm$0.10 & 134$\pm$7 & 11.3$\pm$0.5 & \\
080222B & 501.06$\pm$0.25 & 79.72$\pm$0.27 & 2.244$\pm$0.004 & 13.90$\pm$0.04 & 1405$\pm$13 & 5.363$\pm$0.025 & \\
110222 & 361.6$\pm$3.5 & 16.1$\pm$1.9 & 11.3$\pm$1.7 & 76$\pm$8 & (2.3$\pm$0.7)$\cdot$10$^5$ & <1 & \\
140222B & 78.16$\pm$0.15 & 253.7$\pm$1.5 & 2.917$\pm$0.017 & 5.121$\pm$0.027 & 70.3$\pm$1.1 & <1 & \\
180222 & 1107$\pm$21 & 11.94$\pm$0.23 & 4.70$\pm$0.07 & 25.3$\pm$0.5 & (2.96$\pm$0.16)$\cdot$10$^4$ & 432$\pm$34 & \\
220222 & 283.6$\pm$1.8 & 102.2$\pm$0.6 & 1.690$\pm$0.022 & 17.02$\pm$0.10 & (2.58$\pm$0.05)$\cdot$10$^3$ & 33.9$\pm$1.2 & \\
010322A & 209.4$\pm$1.1 & 387$\pm$6 & 2.13$\pm$0.04 & 10.87$\pm$0.16 & 673$\pm$29 & 2.93$\pm$0.17 & \\
010322B & 17.17$\pm$0.33 & (1.38$\pm$0.28)$\cdot$10$^3$ & 18$\pm$4 & 1.8$\pm$0.4 & 3.1$\pm$1.8 & <1 & \\
080322A & 2.935$\pm$0.019 & 6898$\pm$28 & 3.60$\pm$0.07 & 0.732$\pm$0.006 & 0.205$\pm$0.005 & <1 & \\
080322B & 364.7$\pm$2.2 & 82.9$\pm$0.8 & 1.709$\pm$0.026 & 14.42$\pm$0.16 & (1.57$\pm$0.05)$\cdot$10$^3$ & 25.3$\pm$0.8 & \\
090322B & 25.45$\pm$0.10 & 4831$\pm$29 & 5.974$\pm$0.015 & 0.3274$\pm$0.0013 & 0.01838$\pm$0.00022 & <1 & \\
090322C & (4.185$\pm$0.017)$\cdot$10$^5$ & 255$\pm$4 & 9.11$\pm$0.17 & 7.16$\pm$0.08 & 192$\pm$7 & 106$\pm$20 & \\
100322 & (1.30$\pm$0.14)$\cdot$10$^3$ & 40.3$\pm$3.4 & 1.03$\pm$0.33 & 10.1$\pm$0.8 & (1.9$\pm$0.5)$\cdot$10$^3$ & (1.4$\pm$0.8)$\cdot$10$^2$ & \\
120322 & 19.55$\pm$0.26 & 82$\pm$22 & 140$\pm$34 & 12$\pm$4 & (1.0$\pm$1.1)$\cdot$10$^3$ & <1 & \\
 \hline
\end{tabular}
\end{table*}

\begin{table*}
\scriptsize
\caption{Recorded fireballs with right ascension and declination of the radiant, apparent, geocentric, and heliocentric velocities.}\label{RadVel}
\centering
\begin{tabular}{lcccccccccccc}
\hline
SPMN code &
RA$_{a}$ ($^\circ$) & 
Dec$_{a}$ ($^\circ$) & 
RA$_{g}$ ($^\circ$) & 
Dec$_{g}$ ($^\circ$) &
RA$_{h}$ ($^\circ$) &
Dec$_{h}$ ($^\circ$) &
V$_{a,0}$ (km/s) &
V$_{a,t}$ (km/s) &
V$_{g}$ (km/s) &
V$_{h}$ (km/s) \\
\hline
060222 & 108$\pm$4 & 38.4$\pm$2.2 & 106$\pm$4 & 37.4$\pm$2.5 & 65.6$\pm$0.9 & 5.8$\pm$1.0 & 19.61$\pm$0.09 & 11.229$\pm$0.033 & 16.32$\pm$0.09 & 41.5$\pm$0.5 \\
080222A & 153.14$\pm$0.33 & 2.69$\pm$0.11 & 152.86$\pm$0.34 & 1.70$\pm$0.12 & 106.92$\pm$0.06 & -7.911$\pm$0.031 & 37.17$\pm$0.24 & 16.329$\pm$0.029 & 35.46$\pm$0.25 & 39.91$\pm$0.33 \\
080222B & 135.60$\pm$0.15 & 10.09$\pm$0.04 & 135.18$\pm$0.16 & 8.23$\pm$0.05 & 82.10$\pm$0.05 & -4.635$\pm$0.004 & 23.475$\pm$0.005 & 9.7592$\pm$0.0022 & 20.666$\pm$0.006 & 37.79$\pm$0.04 \\
110222 & 211.58$\pm$0.34 & 71.8$\pm$2.0 & 217.9$\pm$1.1 & 74.3$\pm$2.3 & 60.9$\pm$0.8 & 27.0$\pm$1.7 & 20.1$\pm$0.4 & 15.89$\pm$0.20 & 16.8$\pm$0.4 & 35.24$\pm$0.31 \\
140222B & 41.35$\pm$0.08 & 39.312$\pm$0.028 & 30.63$\pm$0.09 & 35.822$\pm$0.023 & 51.850$\pm$0.018 & 5.641$\pm$0.017 & 15.068$\pm$0.013 & 8.905$\pm$0.004 & 10.536$\pm$0.020 & 39.904$\pm$0.019 \\
180222 & 146.3$\pm$0.5 & 17.90$\pm$0.28 & 145.2$\pm$0.5 & 16.39$\pm$0.29 & 91.44$\pm$0.10 & 1.29$\pm$0.07 & 23.95$\pm$0.08 & 20.045$\pm$0.013 & 21.32$\pm$0.09 & 38.79$\pm$0.23 \\
220222 & 149.508$\pm$0.030 & 30.58$\pm$0.06 & 140.248$\pm$0.028 & 23.63$\pm$0.11 & 80.31$\pm$0.07 & 2.559$\pm$0.022 & 15.60$\pm$0.04 & 5.9502$\pm$0.0031 & 11.41$\pm$0.06 & 35.225$\pm$0.034 \\
010322A & 299.19$\pm$0.11 & 50.410$\pm$0.029 & 304.55$\pm$0.12 & 47.034$\pm$0.012 & 49.777$\pm$0.031 & 34.32$\pm$0.04 & 25.56$\pm$0.05 & 9.799$\pm$0.015 & 22.97$\pm$0.05 & 36.477$\pm$0.009 \\
010322B & 288.24$\pm$0.33 & 54.1$\pm$1.1 & 295.0$\pm$0.4 & 51.9$\pm$1.1 & 56.2$\pm$0.5 & 35$\pm$4 & 23.83$\pm$0.06 & 19.43$\pm$0.13 & 21.01$\pm$0.07 & 34.99$\pm$0.34 \\
080322A & 113.58$\pm$0.07 & -13.672$\pm$0.026 & 104.83$\pm$0.12 & -21.95$\pm$0.07 & 83.91$\pm$0.06 & -13.79$\pm$0.07 & 17.157$\pm$0.030 & 13.560$\pm$0.006 & 13.40$\pm$0.04 & 39.542$\pm$0.029 \\
080322B & 121.84$\pm$0.05 & 10.43$\pm$0.09 & 123.60$\pm$0.04 & 6.96$\pm$0.07 & 91.87$\pm$0.07 & -4.541$\pm$0.026 & 18.72$\pm$0.06 & 8.266$\pm$0.013 & 14.90$\pm$0.08 & 41.49$\pm$0.06 \\
090322B & 259.88$\pm$0.09 & 10.98$\pm$0.09 & 260.24$\pm$0.09 & 10.72$\pm$0.09 & 259.93$\pm$0.27 & 57.545$\pm$0.015 & 63.937$\pm$0.007 & 40.138$\pm$0.033 & 62.749$\pm$0.008 & 41.32$\pm$0.04 \\
090322C & 340.5$\pm$1.2 & 77.58$\pm$0.05 & 6.2$\pm$0.9 & 73.90$\pm$0.14 & 73.418$\pm$0.026 & 18.82$\pm$0.17 & 18.159$\pm$0.018 & 17.96$\pm$0.07 & 14.397$\pm$0.024 & 38.81$\pm$0.05 \\
100322 & 200$\pm$12 & 75.3$\pm$1.7 & 205$\pm$17 & 79.1$\pm$1.8 & 84.42$\pm$0.31 & 24.54$\pm$0.33 & 20.2$\pm$0.7 & 8.358$\pm$0.030 & 17.0$\pm$0.9 & 38.3$\pm$1.4 \\
120322 & 156.87$\pm$0.09 & 28.29$\pm$0.26 & 157.00$\pm$0.10 & 27.05$\pm$0.26 & 104.92$\pm$0.23 & 7$\pm$34 & 21.39$\pm$0.15 & 14.27$\pm$0.06 & 18.27$\pm$0.18 & 40.48$\pm$0.09 \\
 \hline
\end{tabular}
\end{table*}

\begin{figure}
   \centering
   \includegraphics[width=\columnwidth]{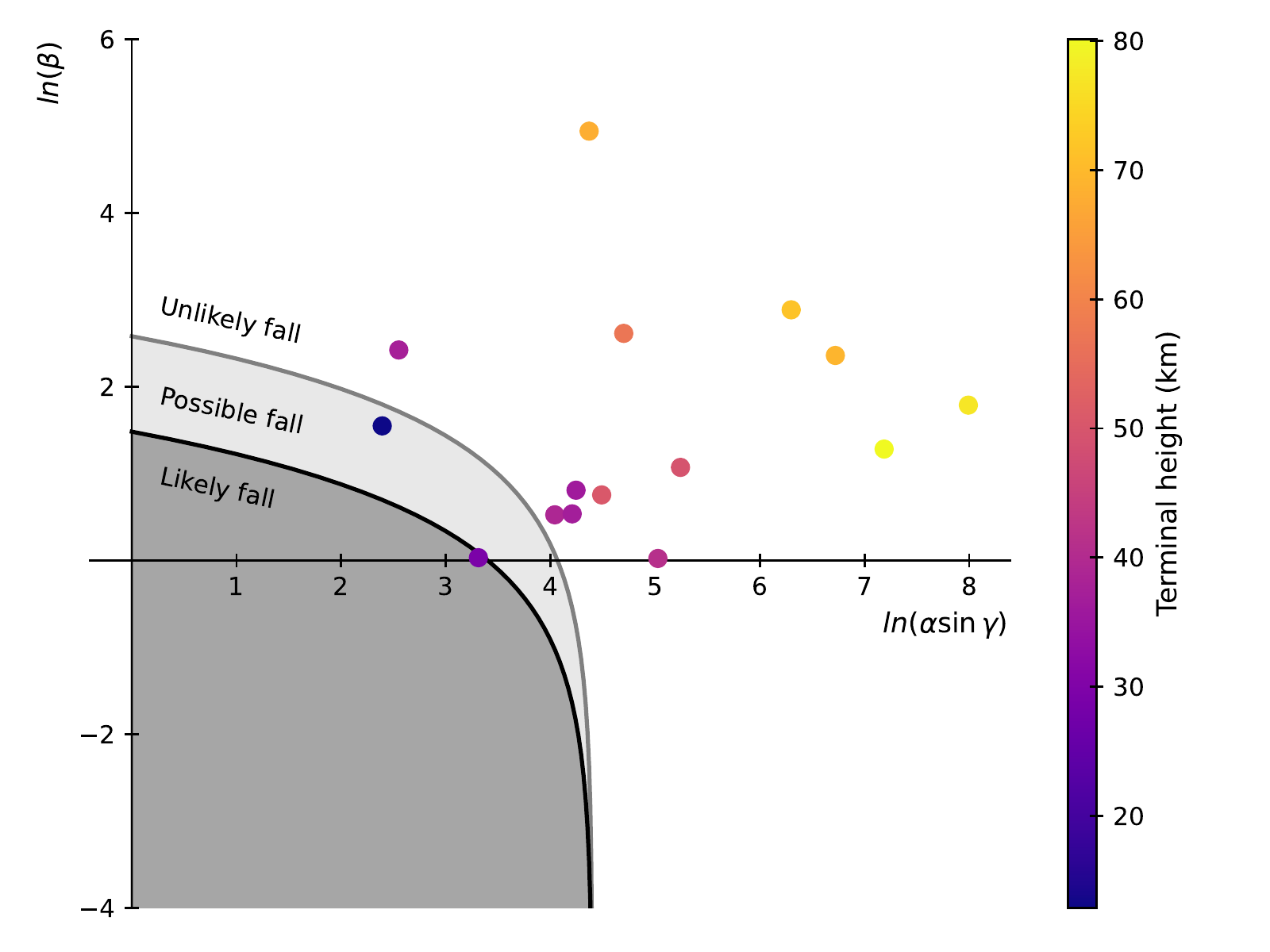}
   \caption{Distribution of the 15 fireballs analyzed over the Spanish territory during February and March 2022 according to the $\alpha-\beta$ criterion. The color bar shows the terminal height, the gray solid curve the boundary for a 50 g meteorite assuming no spin of the meteoroid, and the black solid curve the boundary for a 50 g meteorite assuming equal ablation over the entire meteoroid surface. We assume $\mu=2/3$ for all meteoroids.}
              \label{alpha_beta}%
    \end{figure}

Two meteoroids penetrate up to $\sim$ 30 and 13 km altitude starting the dark flight at a velocity of $\sim$ 8 and 20 km/s, respectively. As can be seen in Figure \ref{alpha_beta}, from the application of the $\alpha-\beta$ criterion and assuming 50 g as the minimum terminal mass to produce a recoverable fall, event SPMN100322 had some possibility of generating a meteorite with a mass of $\sim$140 g, and event SPMN180222 was likely to be a $\sim$430 g meteorite dropper. Unfortunately, a field search campaign was prepared but no fragments were recovered.

\begin{figure}
   \centering
   \includegraphics[width=\columnwidth]{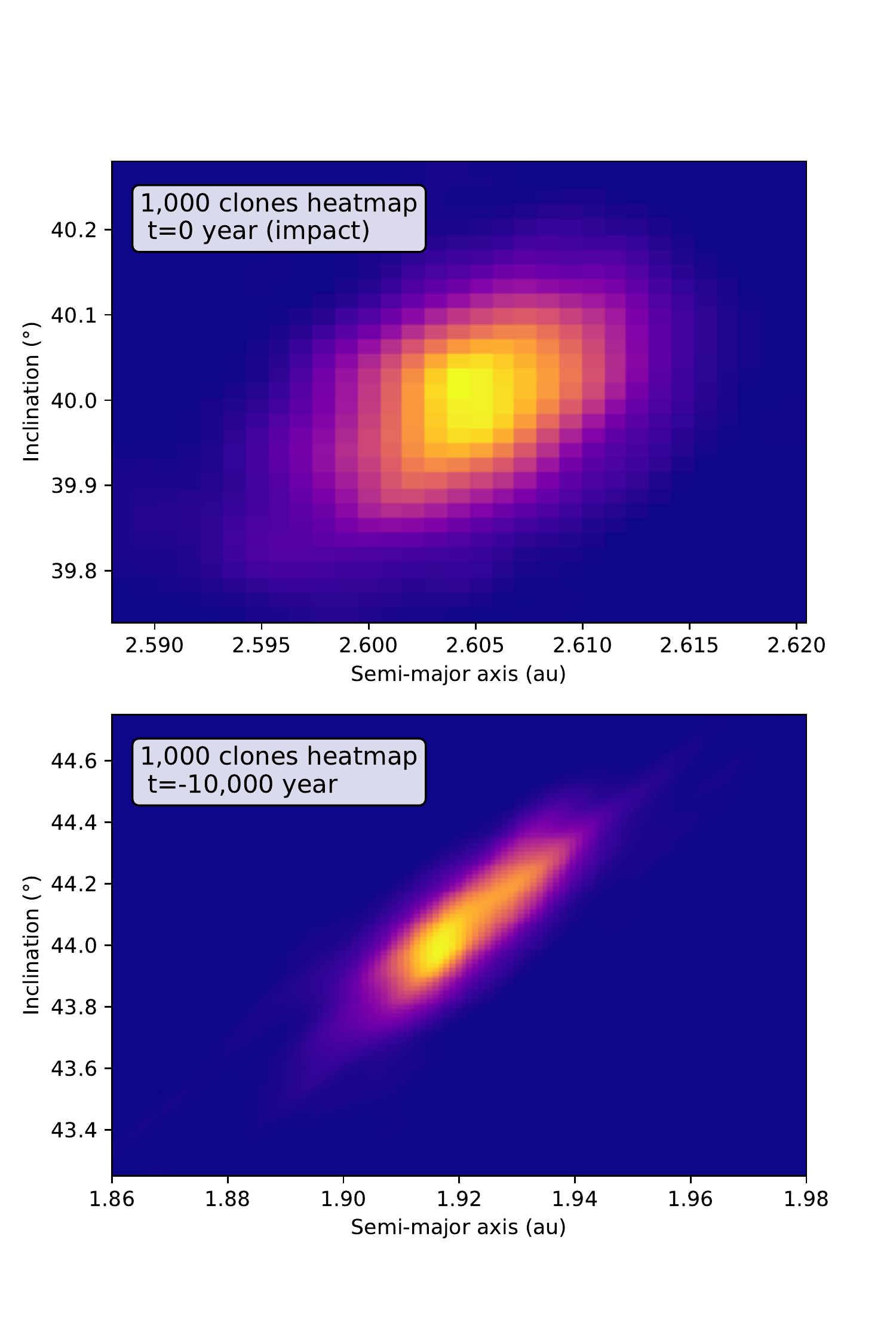}
   \caption{Typical heatmap of the inclination and semi-major axis distribution of the 1,000 clones for the SPMN010322A in the Monte Carlo simulation. The top figure corresponds to the time of impact (t=0 year) without Earth-Moon gravitational focusing correction. The bottom figure corresponds to the end of the backward orbital integration (t=-10,000 years).}
              \label{MC_i_vs_a_SPMN010322A}%
    \end{figure}

The computed osculating orbital elements at the time of impact of the analyzed fireballs are compiled in Table \ref{OrbElem}. As an example of the Monte Carlo simulation, Figure \ref{MC_i_vs_a_SPMN010322A} shows a heat map of the semi-major axis and inclination distribution for the 1,000 clones of event SPMN010322A at the time of impact (t=0 year without Earth-Moon gravitational focusing correction) and at the end of the backward orbital integration (t=-10,000 year). 

Four orbits present very high eccentricity values with large semi-major axes, five can be classified as Jupiter-family comets, while four are asteroid-like orbits. As expected, the orbits tend to be of low inclination, with the exception of SPMN090322B which has an inclination of 122$^\circ$. None of the meteoroids had close encounters with the Moon prior to the impact.

\begin{table*}
\caption{Recorded fireballs with semi-major axis, eccentricity, inclination, perihelion distance, argument of the perihelion, ascending node, and Tisserand parameter (referred to the J2000 equinox). Uncertainty for the ascending node is 0.0001$^\circ$. }\label{OrbElem}
\centering
\footnotesize
\begin{tabular}{lcccccccccccc}
\hline
SPMN code & a (au) & e & i ($^\circ$) & q (au) & $\omega$ ($^\circ$) & $\Omega$  ($^\circ$) & T$_j$ \\
\hline
060222 & 11$\pm$5 & 0.92$\pm$0.04 & 6.1$\pm$1.4 & 0.892$\pm$0.011 & 216.823$\pm$0.011 & 317.8516 & 1.60$\pm$0.21 \\
080222A & 4.3$\pm$0.6 & 0.938$\pm$0.007 & 14.69$\pm$0.10 & 0.2658$\pm$0.0030 & 120.6728$\pm$0.0030 & 138.9337 & 1.77$\pm$0.14 \\
080222B & 2.395$\pm$0.018 & 0.7285$\pm$0.0015 & 5.47$\pm$0.05 & 0.6504$\pm$0.0015 & 78.9636$\pm$0.0015 & 139.8736 & 3.015$\pm$0.014 \\
110222 & 1.60$\pm$0.06 & 0.403$\pm$0.020 & 27.2$\pm$1.0 & 0.954$\pm$0.006 & 208.156$\pm$0.006 & 322.0380 & 4.02$\pm$0.11 \\
140222B & 4.344$\pm$0.033 & 0.7739$\pm$0.0017 & 5.655$\pm$0.010 & 0.98217$\pm$0.00005 & 170.87497$\pm$0.00005 & 325.8715 & 2.320$\pm$0.008 \\
180222 & 3.05$\pm$0.18 & 0.783$\pm$0.011 & 1.53$\pm$0.13 & 0.662$\pm$0.005 & 255.517$\pm$0.005 & 329.0900 & 2.60$\pm$0.09 \\
220222 & 1.605$\pm$0.007 & 0.4698$\pm$0.0026 & 2.68$\pm$0.05 & 0.8510$\pm$0.0005 & 235.6854$\pm$0.0005 & 333.2960 & 4.089$\pm$0.013 \\
010322A & 1.9276$\pm$0.0027 & 0.5469$\pm$0.0006 & 36.06$\pm$0.10 & 0.87346$\pm$0.00008 & 131.68867$\pm$0.00008 & 340.1080 & 3.413$\pm$0.004 \\
010322B & 1.57$\pm$0.06 & 0.411$\pm$0.020 & 35.36$\pm$0.14 & 0.922$\pm$0.007 & 139.445$\pm$0.007 & 340.1481 & 4.00$\pm$0.12 \\
080322A & 3.96$\pm$0.04 & 0.7532$\pm$0.0026 & 13.885$\pm$0.012 & 0.97727$\pm$0.00023 & 15.37103$\pm$0.00023 & 167.1195 & 2.392$\pm$0.012 \\
080322B & 13.5$\pm$1.0 & 0.931$\pm$0.005 & 4.68$\pm$0.04 & 0.9330$\pm$0.0004 & 28.9213$\pm$0.0004 & 167.8948 & 1.570$\pm$0.025 \\
090322B & 11.2$\pm$0.5 & 0.911$\pm$0.004 & 122.44$\pm$0.13 & 0.99250$\pm$0.00009 & 178.06764$\pm$0.00009 & 348.2148 & 1.108$\pm$0.020 \\
090322C & 3.16$\pm$0.04 & 0.688$\pm$0.004 & 18.90$\pm$0.07 & 0.98494$\pm$0.00007 & 168.69848$\pm$0.00007 & 348.2915 & 2.665$\pm$0.019 \\
100322 & 2.8$\pm$0.9 & 0.65$\pm$0.12 & 24.60$\pm$0.17 & 0.98426$\pm$0.00012 & 192.22661$\pm$0.00012 & 349.1653 & 2.8$\pm$0.6 \\
120322 & 6.05$\pm$0.30 & 0.862$\pm$0.008 & 7.92$\pm$0.04 & 0.8339$\pm$0.0023 & 229.2663$\pm$0.0023 & 352.0240 & 1.93$\pm$0.04 \\
 \hline
\end{tabular}
\end{table*}

%--------------------------------------------------------------------
%--------------------------------------------------------------------

%--------------------------------------------------------------------
\section{Dynamic association with meteoroid streams and parent bodies}\label{secDynAsso}
\begin{figure*}
   \centering
   \includegraphics[width=2\columnwidth]{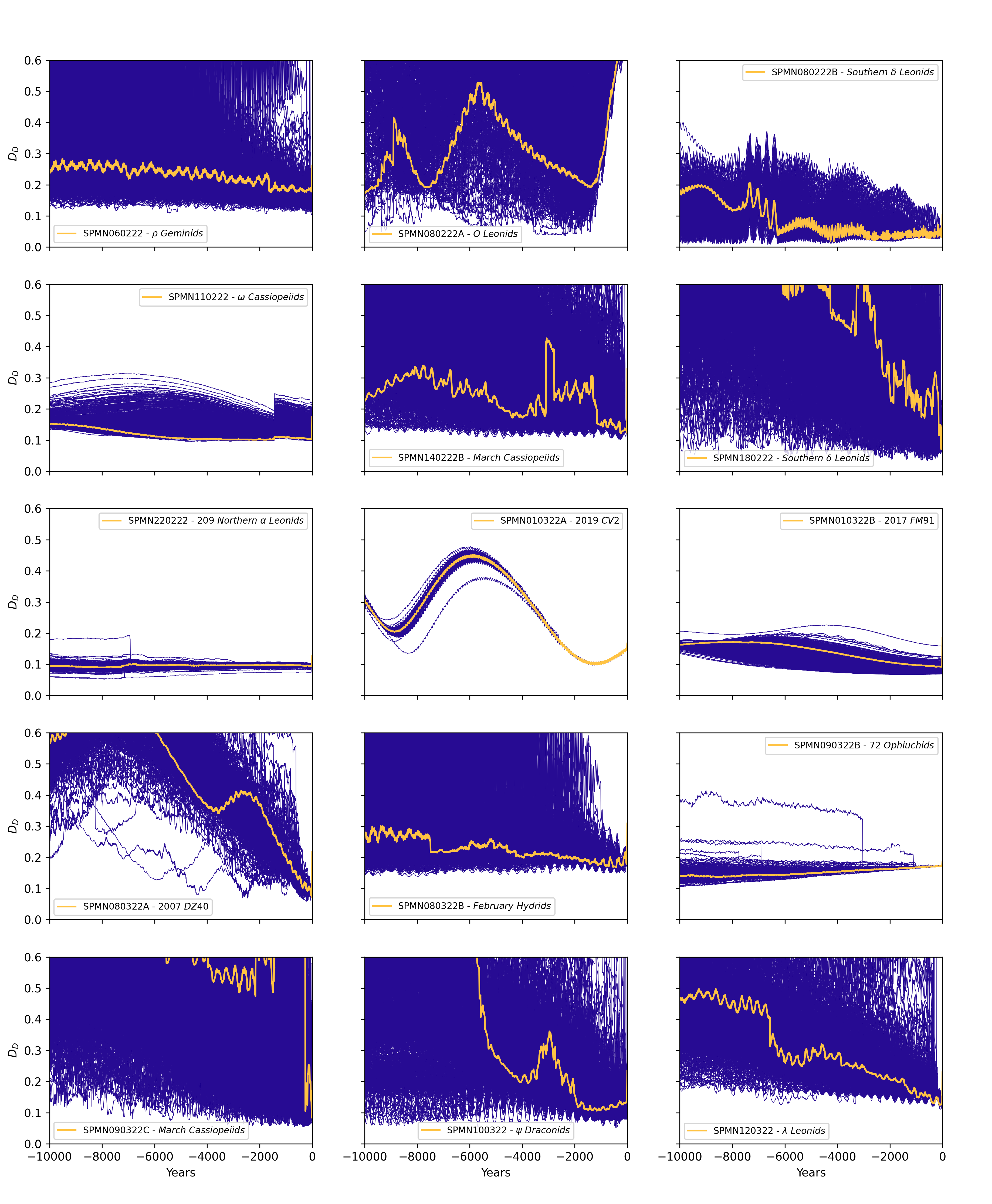}
   \caption{Evolution of the dissimilarity function $D_D$ of the 15 meteoroids with their most favorable candidates during the orbital backward integration over 10,000. The 1,000 clones of each event are also shown. }
              \label{integrations}%
    \end{figure*}

%--------------------------------------------------------------------
The study of the associations of meteoroids that impact our planet with parent bodies or meteoroid streams is not a trivial task. There are numerous mechanisms that prevent the correct linking of meteors with their origins, from the intrinsically chaotic behavior of planetary systems to non-gravitational effects and sporadic collisions and interactions \citep{Trigo2005}. Because of the high probability that two orbits are randomly associated \citep{WiegertBrown2004}, we have not only analyzed the similarity of the orbits at the time of impact but also studied their robustness over time. From the time evolution of the parent body dissimilarity criterion, we found some dynamic associations. Figure \ref{integrations} shows the evolution of the dissimilarity criterion during the orbital integration of the 15 events analyzed in this work, along with their most favorable parent body candidates or meteor shower. Table \ref{refAsso} shows each event with its most likely association, along with the years of time it lasts under the $D_D$ threshold, the minimum encounter distance, the required ejection velocity at the time of minimum distance, and the minimum required ejection velocity. 

5 out of 15 events, that is, about 30\% of the bright fireballs, are below the cut-off for at least 5,000 years. 4 events would be associated with minor showers ($\sim$27\%) and 1 fireball associated with a near-Earth asteroid ($\sim$7\%). In all the associated cases, the required ejection velocity needed to transform the parent orbit into the meteoroid orbit is in good agreement with the estimated range for collisions between objects, which can produce a kick of a few kilometers per second \citep{Melosh1984}.

\begin{table*}
\caption[Stations]{Most likely parent body and meteoroid stream candidates for each event with the minimum $D_D$ value, the years that fulfill the $D_D$ criterion threshold, the minimum encounter distance, the required ejection velocity at the time of minimum distance, and the minimum required ejection velocity during the orbital integration.}
\label{refAsso}
\centering
\begin{tabular}{lcccccc}
\hline
SPMN code & Association & D$_{min}$  & t$_D$ (y) & S$_{min}$ (au) & V$_{S,min}$ (km/s) & V$_{min}$ (km/s) \\
\hline
060222 & $\rho$ Geminids & 0.176 & 180 & 0.186 & 4.7 & 4.7 \\
080222A & o Leonids & 0.174 & 90 & 0.231 & 4.6 & 0.9 \\
080222B & Southern $\delta$ Leonids & 0.018 & 8720 & 0.129 & 0.8 & 0.4 \\
110222 & $\omega$ Cassiopeiids & 0.101 & 10000 & 0.087 & 9.6 & 1.4 \\
140222B & March Cassiopeiids & 0.121 & 1610 & 0.145 & 10.2 & 0.5 \\
180222 & Southern $\delta$ Leonids & 0.07 & 240 & 0.278 & 13.2 & 2.0 \\
220222 & Northern $\alpha$ Leonids & 0.09 & 10000 & 0.05 & 5.8 & 1.4 \\
010322A & 2019 CV2 & 0.099 & 2640 & 0.264 & 6.0 & 1.7 \\
010322B & 2017 FM91 & 0.092 & 9990 & 0.104 & 6.6 & 2.3 \\
080322A & 2007 DZ40 & 0.073 & 800 & 0.144 & 3.1 & 1.1 \\
080322B & February Hydrids & 0.168 & 600 & 0.37 & 15.9 & 3.0 \\
090322B & 72 Ophiuchids & 0.136 & 9990 & 0.811 & 14.3 & 0.4 \\
090322C & March Cassiopeiids & 0.084 & 110 & 0.34 & 10.9 & 0.8 \\
100322 & $\psi$ Draconids & 0.106 & 2080 & 0.37 & 5.1 & 2.0 \\
120322 & $\lambda$ Leonids & 0.125 & 1300 & 0.083 & 7.4 & 2.4 \\
 \hline
\end{tabular}
\end{table*}

%--------------------------------------------------------------------
\section{Discussion}\label{secDisc}

In relation to the various ablation behaviors observed, it is important to note that this could be the result of the differences between chondritic meteoroid and cometary aggregate bulk properties. The low density and high porosity of the latter are directly related to their aerodynamic strengths \citep{Blum2006}. Cometary streams typically produce centimeter-sized projectiles causing fireballs with disruptive flares, and multiple sudden brightness increases or a catastrophic final flare. Due to the heterogeneity of the meteoroid components, the evaporation temperature of each one is reached at different altitudes, giving rise to the so-called differential ablation \citep{Gomez2017}. The aerodynamic overpressure experienced by meteoroids when they fragment allows for estimating their aerodynamic strength. This, in turn, allows for deducing the bulk properties of their meteoroid stream \citep{Wetherill1982, Trigo2006MNRAS}. These types of large fireballs associated with cometary vestiges are the result of rapid disruption in micrometric grains and the sudden ablation of volatile mineral phases driven by the thermal wave in the meteoroid head \citep{Trigo2019}.

Even in such circumstances, it is remarkable that the sporadic contribution is not dominant at all. We found a very significant percentage of bright fireballs dynamically associated with minor showers. Although during the orbital integration there are no very close encounters despite the reasonable ejection velocities, we must point out that we have propagated 18 particles distributed in true anomaly throughout the orbit of the meteoroid streams, but at their nominal values for the rest of the orbital elements. Due to the orbital perturbations accumulated over time and their violent origin, either by tidal forces disruption or catastrophic collisions, the meteoroid streams spread toroidally along their orbit and gradually disperse. Some regions even undergo more pronounced decoherence than others due to the gravitational influence of the Earth-Moon system or nearby planets. 

The minimum ejection velocities calculated to produce the meteoroid orbit from the parent body have a standard deviation range between 0.16 and 1.4 km/s (with an average standard deviation of 0.4 km/s) for the studied events. Although the ejection velocities found are compatible with collisions of small objects in the inner Solar System, this does not necessarily mean that these meteoroids have separated from their meteoroid stream or parent body recently; we just note it as a feasible possibility due to the usual disruption behavior of crumbling asteroids and comets.

Although remarkable, the high number of minor showers producing fireballs should not come as a surprise as such a percentage of meteors associated with meteoroid streams is not unusual. For example, percentages up to 80\% between November and January were already reported belonging to meteor showers \citep{Rao1974}. On the other hand, among the 2,401 records studied by \citet{Lindblad1971}, apparently, 37\% were associated with meteoroid streams. A similar percentage (41\%) was found by \citet{SH1963}. Of the orbits analyzed by \citet{JW1961}, 65\% were linked to a meteor shower. Regarding the Meteorite Observation and Recovery Project (MORP) database, 37\% of the fireballs could be associated with meteoroid stream \citep{Halliday1996}. \citet{Terentjeva1989} performed a grouping according to event candidates to produce meteorites, finding that 68\% of 554 fireballs studied could be part of a shower. And also in good agreement with the results of this work, \citet{Babadjanov1963} reported that of the 185 meteors studied, 73\% appeared to be of cometary origin. Recent studies also show large percentages of meteors associated with meteor showers, for example, 45\% in \citet{Colas2020} and 35\% in \citet{Drolshagen2021}. Regarding superbolides detected from space, 23\% could be associated with meteoroid streams or near-Earth objects \citep{PeAs2022}.

Therefore, as previously studied, it is reasonable to expect that a large percentage of the meteors belong to minor meteoroid streams, but also, as we show in this work, some meteor showers can be a significant source of large projectiles for the Earth and the Moon.

%--------------------------------------------------------------------
\section{Conclusion}\label{secConcl}
%--------------------------------------------------------------------

The extraordinary meteorological conditions in Spain during the spring of 2022 have made it possible to obtain high-quality data related to the fireball activity produced, to a large extent, by minor meteoroid streams. Ground-based multi-station recordings were possible thanks to the ever-increasing atmospheric volume monitored by the SPMN network throughout Spain. We reported 15 bright bolides in February and March, two of them being potential meteorite dropper events. By applying novel computer vision techniques and improved methods of trajectory reconstruction and heliocentric orbit calculation implemented in our software \textit{3D-FireTOC}, we have been able to study in detail the atmospheric flight and dynamic association of large cometary and asteroidal projectiles impacting our planet. Based on the trajectory data, we computed the initial and terminal mass, the aerodynamic strength, and the bulk density by means of an ablation model. In consequence, we claim that:
\begin{itemize}
    \item Among the 169 bright meteors recorded during the spring of 2022 in Spain, 2 of them were potentially meteorite dropper events.
    \item We identify the minor showers o Leonids, Southern $\delta$ Leonids, $\omega$ Cassiopeiids, Northern $\alpha$ Leonids, and 72 Ophiuchids, and the asteroid 2017 FM91 as sources of large projectiles during February and March.
    \item Nearby meteoroid streams can be efficient producers of large projectiles as they account for the $\sim$27\% of the fireballs. 
    \item Near-Earth objects may be a greater source of impact risk than previously thought. 
    \item It is needed to extend the study and cataloguing of minor showers, since, although they are not very active in terms of the number of meteors, our work indicates that they also produce large bolides annually.
    \item These findings support the idea that certain meteoroid streams associated with comets or asteroids may represent a short-term impact hazard. 
\end{itemize}

Finally, we think that understanding the origin and mechanisms by which large meteoroids reach the Earth is of great scientific interest due to the possibility of associating complexes and parent bodies with fireballs and, ultimately, meteorites found on Earth and the Moon. The relevance of associations also reverts in outreach, as we can quickly inform the public about the origin of the fireballs reported by eyewitnesses.

%%%%%%%%%%%%%%%%%%%%%%%%%%%%%%%%%%%%%%%%%%%%%%%%%%

\section*{acknowledgements}
      This project has received funding from the European Research Council (ERC) under the European Union’s Horizon 2020 research and innovation programme (grant agreement No. 865657) for the project “Quantum Chemistry on Interstellar Grains” (QUANTUMGRAIN). JMT-R and E.P-A. acknowledge financial support from project PID2021-128062NB-I00 funded by MCIN/AEI/10.13039/501100011033. AR acknowledge financial support from the FEDER/Ministerio de Ciencia e Innovación – Agencia Estatal de Investigación (PID2021-126427NB-I00, PI: AR). AR is indebted to DIUE (project 2017SGR1323). Cebreros \#AMS81 ESA Ground station belongs to the AllSky7 fireball monitoring project and is operated by Rainer Kresken and Pablo Ramirez Moreta. We also thank all SPMN station operators whose continuous dedication have allowed to record these bolides from multiple stations: Jordi Donet Donet, Vicent Ibáñez, Jose M. Serna, Carlos Alcaraz, Antonio J. Robles, Ramón López, Agustín Núñez, José A. de los Reyes, Sensi Pastor, Antonio Fernández Sánchez, Antonio Lasala, Álex Gómez, Juan Gómez, Ramón López, Francisco José García Rodríguez and Cesar Guasch Besalduch.

\section*{Data Availability}

The data underlying this article will be shared on reasonable request to the corresponding author.



\bibliographystyle{mnras}
\bibliography{example} % if your bibtex file is called example.bib

%\begin{thebibliography}{99}
%\end{thebibliography}

%%%%%%%%%%%%%%%%%%%%%%%%%%%%%%%%%%%%%%%%%%%%%%%%%%

%%%%%%%%%%%%%%%%% APPENDICES %%%%%%%%%%%%%%%%%%%%%

%%%%%%%%%%%%%%%%%%%%%%%%%%%%%%%%%%%%%%%%%%%%%%%%%%

% Don't change these lines
\bsp	% typesetting comment
\label{lastpage}
\end{document}